# Compact low-half-wave-voltage thin film lithium niobate electro-optic phase modulator fabricated by photolithography assisted chemo-mechanical etching

LANG GAO,[1,2] YOUTING LIANG,[3] JINMING CHEN,[3] JIANPING YU,[3] JIA QI,[3] LVBIN SONG,[3] JIAN LIU,[3] ZHAOXIANG LIU,[3] HONGXIN QI,[3,4,*] AND YA CHENG,[1,3,4,5,6,7,8,*]

[1]State Key Laboratory of High Field Laser Physics and CAS Center for Excellence in Ultra-intense Laser Science, Shanghai Institute of Optics and Fine Mechanics (SIOM), Chinese Academy of Sciences (CAS), Shanghai 201800, China
[2]Center of Materials Science and Optoelectronics Engineering, University of Chinese Academy of Sciences, Beijing 100049, China
[3]The Extreme Optoelectromechanics Laboratory (XXL), School of Physics and Electronic Science, East China Normal University, Shanghai 200241, China
[4]State Key Laboratory of Precision Spectroscopy, East China Normal University, Shanghai 200062, China
[5]Collaborative Innovation Center of Extreme Optics, Shanxi University, Taiyuan, Shanxi 030006, People's Republic of China
[6]Collaborative Innovation Center of Light Manipulations and Applications, Shandong Normal University, Jinan 250358, People's Republic of China
[7]Hefei National Laboratory, Hefei 230088, China
[8]Joint Research Center of Light Manipulation Science and Photonic Integrated Chip of East China Normal University and Shandong Normal University, East China Normal University, Shanghai 200241, China
*hxqi@phy.ecnu.edu.cn
*ya.cheng@siom.ac.cn

**This paper presents a compact dual-arm thin film lithium niobate (TFLN) electro-optic phase modulator fabricated using the photolithography-assisted chemo-mechanical etching (PLACE) technique. The design of the device allows for complete utilization of the microwave electric field, doubling the modulation efficiency compared to single-arm modulators in theory. With a half-wave voltage of approximately 3 V and a modulation length of 1 cm, the device outperforms conventional phase modulators. Furthermore, the phase modulator exhibits low sensitivity to optical wavelengths in the range of 1510-1600 nm and offers a low insertion loss of 2.8 dB. The capability to generate multiple sideband signals for optical frequency comb applications is also demonstrated, producing 29 sideband signals at an input microwave power of 2 W.**

Phase modulator is one of the essential functional elements in photonic integrated circuits and microwave photonic systems[1], which can be applied to optical communication[2], sensing[3], frequency combs[2, 4], and radio-over-fiber systems[1]. Currently, integrated phase modulators based on various materials have been investigated, including silicon-based[3, 5, 6], indium phosphide (InP)[7, 8], aluminum nitride (AlN)[9], graphene[10, 11], polymeric materials[12], and lithium niobate[13-15]. Among them, the electro-optic phase modulator built on lithium niobate platform has been most widely investigated due to its superior material properties, including a superior linear electro-optic coefficient ($\gamma_{33}$), thermal stability, and a wider transparent window range[13, 14].

Conventional lithium niobate phase modulators are built on bulk material and use the Ti diffusion process to fabricate the optical waveguides. The waveguides fabricated in this way have low refractive index contrast ($\Delta n \approx 0.01$) and large optical mode field, resulting in low modulation efficiency, high half-wave voltage ($V_\pi \approx 10\ V$), and large device footprint[13, 14]. Recently, remarkable progress has been achieved in the fabrication of thin-film lithium niobite (TFLN) wafers; electro-optic phase modulators based on TFLN are expected to improve modulation efficiency further and reduce half-wave voltage while decreasing device footprint in the meantime. Some phase modulators based on TFLN have been reported with a half-wave voltage as low as ~4.5 V and a waveguide length more than 2 cm[16]. However, these phase modulators are all composed of a pair of ground-signal-ground (GSG) electrodes, where only one gap of GSG electrodes is used, which means that only half of the microwave electric field takes effect in modulation of the optical phase in the waveguide.

In this paper, we demonstrate a compact dual-arm TFLN phase modulator fabricated by photolithography assisted chemo-mechanical etching (PLACE) technique. Through unique design, full utilization of the microwave electric field is achieved, which allows for doubling the modulation efficiency in theory. We successfully fabricated the device using the PLACE technology and evaluated its half-wave voltage, generation of sideband signals, and insertion loss performance. Our device exhibits a low insert loss of 2.8 dB thanks to the efficient spot size converters and low-propagation-loss waveguides.

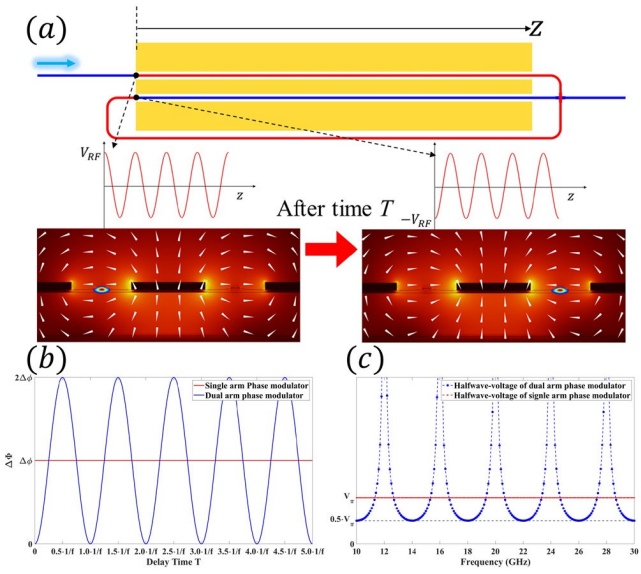

**Fig. 1.** (a) Schematic illustration of the dual-arm Thin Film Lithium Niobate (TFLN) phase modulator principle. (b) Phase modulation amount as a function of delay time T. (c) Simulated half-wave voltage $V_\pi$ as a function of microwave frequency $f$.

The schematic diagram of the dual-arm TFLN phase modulator is shown in Fig. 1(a). Consider a light signal with an amplitude of $E_0$ and an angular frequency of $\omega_0$, denoted as $E_{in} = E_0 e^{j\omega_0 t}$, is injected into the phase modulator; the output signal after phase modulation can be expressed as $E_{out} = E_0 e^{j\omega_0 t} e^{j\Delta\phi}$, where $\Delta\phi$ is the phase variation under the electro-optic modulation, which can be expressed as $\Delta\phi = \pi \frac{V_{RF}}{V_\pi} - \pi \frac{V_{RF}}{V_\pi} \cos(2\pi f T)$, where $V_\pi$, $V_{RF}$, and $f$ represent half-wave voltages of the single-arm phase modulator, peak voltage of microwave signal, and frequency of microwave signal respectively, $T$ represents the delay time for light signal to pass through the delay waveguide (represented by the red waveguides) as shown in Fig. 1(b). When $T = m \cdot (1/f)$ ($m = 1,2,3,......$), the total amount of phase modulation of output signal is 0. This is because the direction of the electric field between the upper and lower electrode pairs are opposite; whereas after the delay time $T$, the electric field between the paired electrodes does not flip over, the light signal experiences an opposite modulation effect in the gap below the electrodes. Finally, the phase modulations obtained in the upper and lower electrode pairs cancel out, which, overall, shows no modulation. While at times $T = (m + \frac{1}{2}) \cdot (1/f)$ ($m = 1,2,3,......$), $\Delta\phi = 2 \cdot \pi \cdot V_{RF}/V_\pi$, the total amount of phase modulation of output light signal is twice that of single-arm modulation. This is because after delay time $T$, the electric field between electrodes flips over, and the light signal experiences same modulation in the electrode pairs, showing that the modulation amount becomes twice that of single-arm modulation. The variation relationship of phase modulation amount with delay time $T$ is shown in Fig. 1(b), where the phase modulation amount increases twice when the delay time $T = 0.5 \cdot (1/f)$, $1.5 \cdot (1/f)$, ...... We also simulated the relationship of the $V_\pi$ with microwave frequency $f$, as shown in Fig. 1(c). From Fig. 1(c), it can be observed that $V_\pi$ exhibits a periodic variation pattern with frequency. $V_\pi$ decreases by half at some specific frequencies, corresponding to the conditions which leads to the doubling of the phase modulation efficiency. While at other frequencies, $V_\pi$ sharply increases, corresponding to the frequencies at which the phase modulation amount is zero.

The design of the dual-arm TFLN phase modulator is shown in Fig. 2(a), and the microscope image of the fabricated phase modulator is shown in Fig. 2(b). The phase modulator mainly consists of three parts, including low-loss delay waveguides and cross waveguides prepared by PLACE technology, a pair of coplanar waveguides traveling wave electrodes with a total electrode length of 1 cm, and two spot size converters (SSC) located at both ends of the device. In Fig. 2(c), we present the cross-sectional view of the device corresponding to the phase shifter section. The low-loss optical waveguide features an TFLN ridge waveguide with a top width of 1 μm, a bottom width of 4.8 μm, and an etching depth of 210 nm. The signal electrode has a width of 23.5 μm, and the gap between the electrodes on the two sides of the waveguide is 6.5 μm. This gap is designed to be as small as possible to achieve a low half-wave voltage while ensuring reasonably low propagation loss for the light traveling within the waveguide. It is crucial that the two electrodes are sufficiently far away from the optical mode in the waveguide. Figure 2(d) presents the simulated optical and electrical field contour plots. At both ends of the electrodes, taper regions are designed to match the dimensions of the high-frequency probe, as present in Figure 2(b). The width of the signal electrode expands from 23.5 μm to 60 μm, and the electrode gap expands from 6.5 μm to 20 μm. This design ensures good impedance matching with a 50 Ω load. The contour plot of the optical fields at the waveguide cross is shown in Fig. 2(e).

The device was fabricated on a 500-nm-thick X-cut TFLN, bonded to a buried oxide (SiO$_2$) layer of thickness 4.7 μm. The optical waveguide is fabricated using the PLACE technique. In this process, a chrome mask was produced on the TFLN substrate by femtosecond laser surface texturing. The mask pattern was then transferred to the TFLN substrate by chemical mechanical polishing.

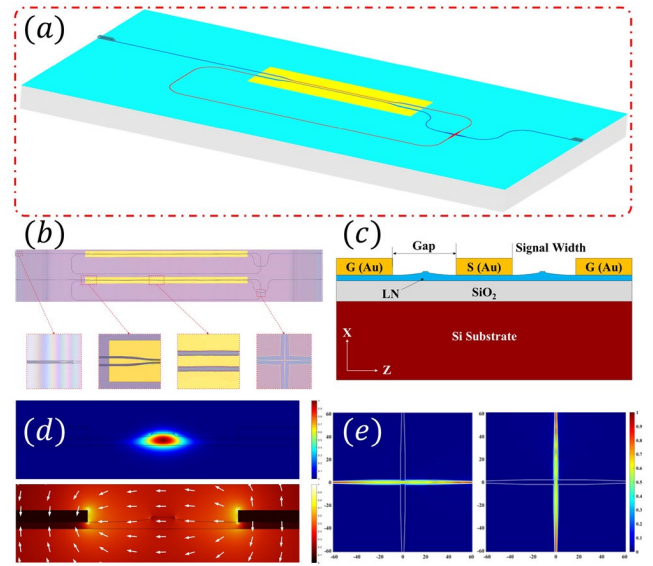

**Fig. 2.** (a) Schematic diagram of a dual-arm TFLN phase modulator. (b) Optical micrograph of the dual-arm TFLN phase modulator chip. (c) Cross-sectional structure of the phase shifter area. (d) Simulated optical field and electrical field. (e) Simulated optical field in the cross waveguide.

More detailed information can be found in Ref. [17,18]. The ground-signal-ground (GSG) coplanar waveguide electrodes were fabricated using femtosecond laser photolithography along with chemical wet etching methods[19]. The fabrication procedure for the electrodes consisted of several steps: (1) deposition of a 500-nm-thick gold (Au) layer by electron beam evaporation, (2) deposition of 200-nm-thick titanium (Ti) layer by magnetron sputtering, (3) patterning of the Ti layer by femtosecond laser ablation, (4) wet etching of the unprotected area of the Au layer by aqua regia, and (5) removal of the Ti mask by sulfuric acid. These steps successfully fabricated the desired GSG electrodes. In the final step of device fabrication, we first formed an LN waveguide taper and covered it with a silicon oxynitride (SiON) waveguide with a cross-section size of 4 μm×3 μm, thus fabricating an optical spot size converter (SSC).

The fabricated dual-arm TFLN phase modulator are characterized by the experimental setup shown in Fig. 3. The tunable optical signal from the polarization controller was coupled into the spot-size converter at one end of the waveguide, using a 3-μm-mode-field-diameter fiber. The modulated signals were then coupled back into the same type of fiber through the spot size converter. The output signal was further directed to an optical spectral analyzer (Yokogawa AQ6370D) for analysis. The modulation signals from the microwave source were injected into the microelectrodes through a ground-signal-ground (GSG) probe, and the other end of the microelectrodes was connected to a 50 Ω matching load. The modulated output signal can be expressed as $E_{out} = E_0 e^{j\omega_0 t} e^{j\Delta\phi}$, where $\Delta\phi = \pi \frac{V_{RF}}{V_\pi} \sin(\omega_f t)$ is the dynamic phase modulation. The output signal is expanded by the first kind of Bessel function (i.e., $J_n(m), n = 0,1,...$) as

$$\mathrm{Re}(E_{out}) = E_0 J_0(m)\cos(\omega_0 t)$$
$$+ E_0 \sum_{n=1}^{\infty} J_n(m)\cos[(\omega_0 + n\omega_f)t] \quad (1)$$
$$+ E_0 \sum_{n=1}^{\infty} (-1)^n J_n(m)\cos[(\omega_0 - n\omega_f)t]$$

Where the intensity ratio of zero-order signal to first-order signal is $J_0^2(m)/J_1^2(m)$, which is a function of modulation amount $m$ ($m = \pi V_{RF}/V_\pi$). The extracted intensity ratio of zero-order signal to first-order signal from OSA can be converted into half-wave voltage $V_\pi$ corresponding to the phase modulator.

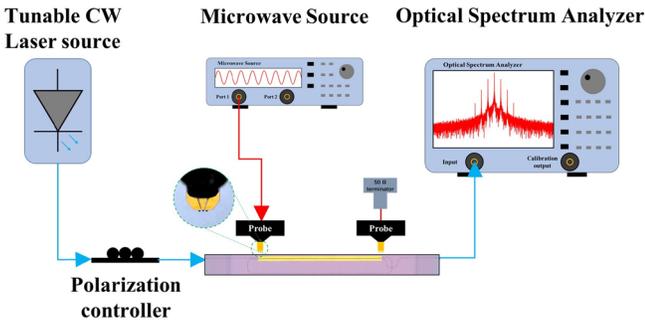

**Fig. 3.** Measurement setup for the dual-arm TFLN phase modulator.

Figure 4(a) shows the relationship of half-wave voltage with microwave frequency for the phase modulator. Near the frequencies of 12 GHz, 16 GHz, 22 GHz, and 26 GHz the measured half-wave voltage is about 3 V, which is about half of that of a single-arm phase modulator ($V_\pi = 5.5$ V), as shown in the red line in

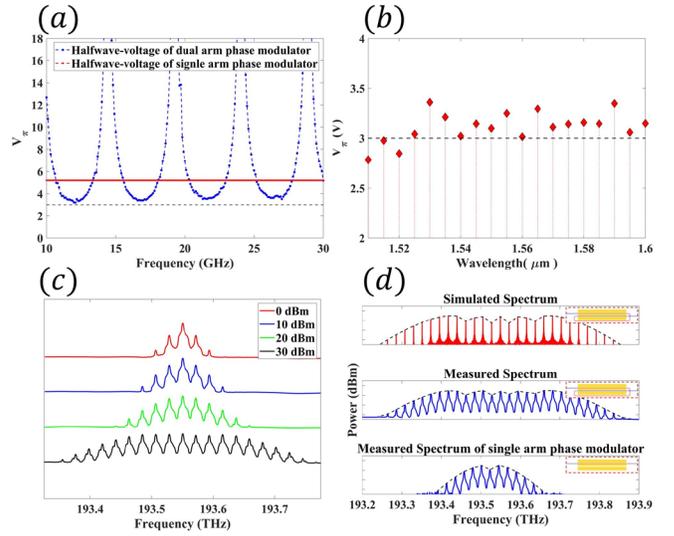

**Fig. 4.** (a) Half-wave voltage as a function of microwave frequency $f$. (b) The half-wave voltage versus optical wavelength. (c) The optical spectra for different microwave input powers. (d) Comparison of different output spectra: simulated, measured, and single-arm phase modulator spectra.

Fig 4(a). The relationship is consistent with the simulation results shown in Fig. 1(c). It is worth noting that there is a certain deviation in frequency between the measured data and the simulated data, which is caused by a fabrication error in the length of waveguide. This deviation can be compensated for through a simple adjustable delay line[19] to match any microwave frequency. Fig 4(b) shows the sensitivity of half-wave voltage to optical wavelength, in the range of 1510 nm – 1600 nm, the $V_\pi$ remains at around 3 V. This insensitivity of half-wave voltage to wavelength is favorable for integration with arbitrary wavelength on-chip lasers. And the fluctuation of $V_\pi$ in the measured data may come from jitter noise of microwave signal source.

An important application of phase modulators is electro-optic frequency comb generation, which has received much attention in the fields of communications, ultrafast optics and so on. The main challenge for realizing an electro-optic frequency comb with a single phase modulator is the microwave power consumption, which requires lower half-wave voltages. Fig 4(c) shows the generation of sideband signals under different microwave power inputs. Fig 4(d) shows the spectral lines of the modulated output signal at about 2 W microwave power input, and the spectral shape is consistent with theoretical simulation. In our experiment, about 13 sideband signals are generated by a single arm phase modulator at about 2 W power, whereas about 29 sideband signals are generated by the dual-arm phase modulator. The results clearly show that the modulation efficiency of the dual-arm TFLN phase modulator is much higher than that of single arm phase modulator.

In summary, leveraging the self-developed PLACE fabrication technology, we demonstrate a compact dual-arm TFLN phase modulator featuring low half-wave voltage and low insertion loss. This innovation effectively addresses the shortcomings of large half-wave voltage, large structural size, and insufficient consumption of microwave power in the electro-optic modulation. Specifically, a half-wave voltage $V_\pi$ of 3 V is achieved with a modulation length of 1 cm. Furthermore, for the application of the phase modulator in the generation of optical frequency combs, we demonstrate its capability to generate a large number of sideband signals. In fact, under a 2 W microwave power input, 29 sideband signals are produced. Our work significantly contributes to the development of high-performance integrated electro-optical combs, as well as low-power consumption communication and sensing systems based on phase modulators.

**Funding.** National Key R&D Program of China (2019YFA0705000), National Natural Science Foundation of China (Grant Nos. 12192251, 12334014, 11933005, 12134001, 61991444, 12204176, 12274133), Science and Technology Commission of Shanghai Municipality (NO.21DZ1101500). Innovation Program for Quantum Science and Technology (2021ZD0301403).

**Disclosures**. The authors declare no conflict of interest.

**Data Availability.** Data underlying the results presented in this paper are not publicly available at this time but may be obtained from the authors upon reasonable request.